\documentclass[aps,prx, reprint,superscriptaddress, longbibiography]{revtex4-1}
\usepackage{color}
\usepackage{soul}
\usepackage{dcolumn}
\usepackage{graphicx}
\usepackage{amsmath}
\usepackage{amssymb}
\usepackage{bm}

\begin{document}










\title{The high-coherence fluxonium qubit}


\author{Long B. Nguyen}
\affiliation{Department of Physics, Joint Quantum Institute, and Center for Nanophysics and Advanced Materials,
	University of Maryland, College Park, MD 20742, USA.}

\author{Yen-Hsiang Lin}
\affiliation{Department of Physics, Joint Quantum Institute, and Center for Nanophysics and Advanced Materials,
	University of Maryland, College Park, MD 20742, USA.}

\author{Aaron Somoroff}
\affiliation{Department of Physics, Joint Quantum Institute, and Center for Nanophysics and Advanced Materials,
	University of Maryland, College Park, MD 20742, USA.}

\author{Ray Mencia}
\affiliation{Department of Physics, Joint Quantum Institute, and Center for Nanophysics and Advanced Materials,
	University of Maryland, College Park, MD 20742, USA.}

\author{Nicholas Grabon}
\affiliation{Department of Physics, Joint Quantum Institute, and Center for Nanophysics and Advanced Materials,
	University of Maryland, College Park, MD 20742, USA.}

\author{Vladimir E. Manucharyan}
\affiliation{Department of Physics, Joint Quantum Institute, and Center for Nanophysics and Advanced Materials,
University of Maryland, College Park, MD 20742, USA.}



\date{\today}

\begin{abstract}


We report superconducting fluxonium qubits with coherence times largely limited by energy relaxation and reproducibly satisfying  $T_2 > 100~\mu s$ ($T_2 > 300~\mu s$ in one device). 
Moreover, given the state of the art values of the surface loss tangent and the $1/f$ flux noise amplitude, coherence can be further improved beyond $1~\textrm{ms}$. Our results violate a common viewpoint that the number of Josephson junctions in a superconducting circuit -- over $10^2$ here -- must be minimized for best qubit coherence. We outline how the unique to fluxonium combination of long coherence time and large anharmonicity can benefit both gate-based and adiabatic quantum computing.\\

\end{abstract}

\maketitle

\section{Introduction}


%

Quantum superconducting circuits based on Josephson tunnel junctions have become a leading platform in the pursuit of quantum computing~\cite{devoret2013superconducting}. These artificial ``atoms" can be printed on a chip in large numbers, wired together for strong interactions, and precisely manipulated and read by RF electronics~\cite{schoelkopf2008wiring}. The Josephson tunnel junction provides the necessary non-dissipative non-linearity required to turn linear electrical circuits into quantum bits (qubits) and strong circuit-circuit coupling into fast logical operations. The weak point of this platform is the relatively short coherence time of physical qubits, as compared to conventional atomic systems~\cite{ladd2010quantum}. It introduces errors during the gate operations~\cite{barends2014superconducting} and constrains the number of qubits that can coherently tunnel in a quantum annealer~\cite{boixo2016computational}. With the growing interest in complex superconducting quantum processors~\cite{kandala2017hardware, versluis2017scalable, PhysRevLett.119.180511,otterbach2017unsupervised, neill2018blueprint}, improving coherence of physical qubits without sacrificing their controllability remains a central problem.  

Material imperfections, in the form of the dielectric loss and the $1/f$ flux noise, are the major obstacles for extending coherence of superconducitng qubits. The former effect is due to microscopic two-level charge defects residing in the interface oxide layers of a typical thin film device~\cite{martinis2005decoherence}. As a consequence, each circuit capacitance acquires a non-zero loss tangent which induces the energy relaxation of the qubit. The latter effect is due to the unpaired electrons trapped in the same oxide layers and acting as the spin-$1/2$ impurities. Their low-temperature dynamics generates a noisy magnetic flux through any superconducting loop with a $1/f$ type spectral density~\cite{koch2007model}. Hence, flux-tuning of qubits generally comes at the expense of reduced coherence time. In the case of conventional flux qubits, coherence time rapidly drops to a few nanoseconds upon detuning from the half-integer flux bias~\cite{yoshihara2006decoherence, orgiazzi2016flux}. Furthermore, recent studies suggest that flux noise can induce a rapid energy relaxation of flux qubits, presumably through the absorption of $\textrm{GHz}$-frequency photons by the spin defects~\cite{yan2016flux, quintana2017observation}. 

Upgrading circuit materials proved a challenging task~\cite{kline2011sub,kumar2016origin}. Alternatively, coherence can be improved by designing noise-insensitive circuits. The successful tricks so far all sacrificed the qubit anharmonicity. For example, the transmon qubit is derived from a Cooper pair box by shunting the junction with a large external capacitance~\cite{koch2007charge}. The qubit sensitivity to the $1/f$ charge noise dropped exponentially, but the spectrum of the circuit also evolved from that of a nearly perfect two-level system to that of an oscillator with about $5\%$ anharmonicity. Capacitive shunting of a flux qubit helped to reduce its unnecessarily large flux sensitivity, but with a similar reduction of anharmonicity~\cite{steffen2010high}. The low anharmonicity can present novel challenges for scaling. One generic problem is that the non-linearity enables logic, and hence reducing it too much will inevitably slow down the gates. For example, the dispersive shift in circuit quantum electrodynamics (cQED) drops with the qubit-oscillator detuning $\Delta$ much faster for a transmon ($1/\Delta^2$) than for the charge qubit ($1/\Delta$)~\cite{koch2007charge}. Another problem is that exciting multiple qubits creates nearly resonant conditions for state leakage outside of the computational subspace, a serious error which is difficult to correct algorithmically.

\begin{figure*}
	\centering
	\includegraphics[width=0.7\linewidth]{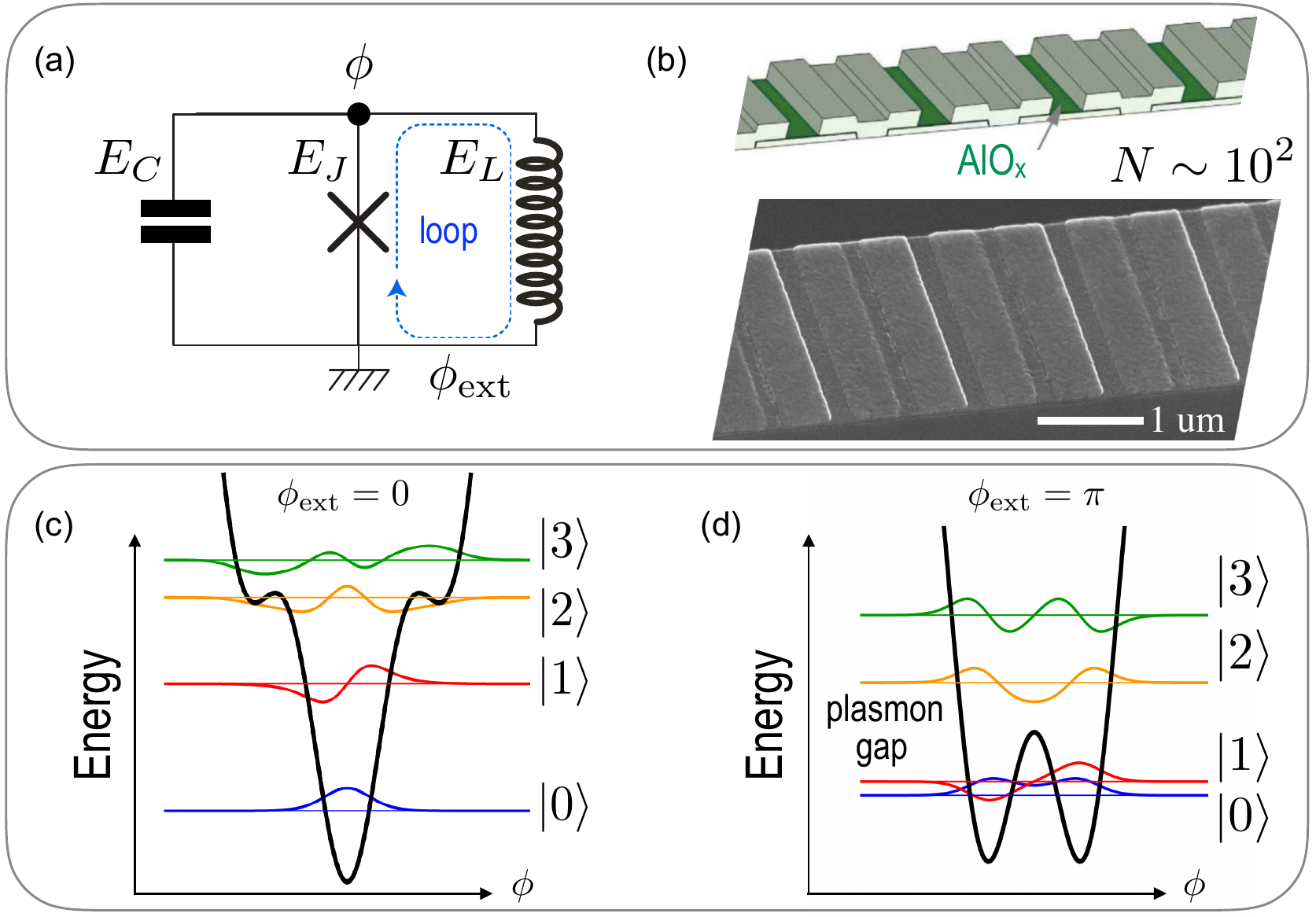}
	\caption{
		(a) The three-element circuit model of fluxonium. (b) Implementation of the large-value inductance $L$ using a linear chain of properly chosen Josephson junctions. (c) The particle-in-a-box potential profile, the spectrum, and the eigenstates at an integer flux through the loop. The $0-1$ qubit transition is qualitatively similar to a transmon. (d) Same as (c) at a half-integer flux bias. The tunnel split qubit states are separated from the non-computational states by a gap associated with exciting the plasma-like oscillations of the phase $\phi$.   
	}
	
	\label{fig:Fig1}
\end{figure*} 

Topological states of matter can in principle provide a foundation for the ultimate protected qubit with a practically infinite coherence time~\cite{kitaev2003fault}. Unfortunately, even for the simplest topological protection scenarios, outlined for the ``$0-\pi$"-type qubits, theory places challenging, if not conflicting, requirements on circuit parameters~\cite{ioffe2002possible, brooks2013protected}. In addition, the existence of protected gate operations with such devices remains unclear~\cite{groszkowski2018coherence}. A more practical quantum memory can be achieved by storing a qubit using non-classical states of radiation in high quality factor linear resonators. This is motivated by the availability of superconducting cavities with a photon loss rate well into the $\textrm{ms}$ regime and no appreciable dephasing mechanisms~\cite{reagor2013reaching}. Non-tunable transmons are used solely as microwave-activated switches for performing gate operations~\cite{wang2016schrodinger}. 
However, oscillator-based encoding is tailored for specific experiments and cannot be straightforwardly applied to advance other directions in superconducting or hybrid quantum computing. Therefore there is still a large demand for a versatile high-performance superconducting qubit.

In this work we describe \textit{fluxonium} superconducting qubits~\cite{Manucharyan113} designed to evade both the surface loss and the noise induced decoherence, without sacrificing the anharmonicity, the flux-tuning range, or the controllable interactions. The present design~\cite{lin2018demonstration} features a capacitive antenna directly connected to the weak junction for compatibility with the widely used transmon-type qubits. We observed $T_2 > 100~\mu s$ at the half-integer flux bias in 7 out of 8 devices with varying circuit parameters. One device had $T_2 > 200~\mu s$, another device had $T_2 > 300~\mu s$, while the single outlier device had $T_2 \approx 80-90~\mu s$. These numbers were reproducible after thermal cycles. Analyzing the data from multiple devices, we conclude that coherence is still largely limited by the surface loss in the antenna because of a suboptimal aluminum-on-silicon fabrication procedure.

%



\subsection{Fluxonium}


A fluxonium circuit consists of a Josephson junction with energy $E_J$  shunted by a capacitance $C$ and an inductance $L$. The two linear elements introduce the charging energy $E_C = e^2/2C$ and the inductive energy $E_L = (\hbar/2e)^2/L$. The parameters must satisfy $E_L \ll E_J$ and $1 \lesssim E_J/E_C \lesssim 10$, which distinguish fluxonium from other inductively-shunted junction devices. These conditions place a challenging requirement on the value of $E_L$, which translates to an extremely large inductance per unit length of about  $10^4 \mu_0$, where $\mu_0$ is the vacuum permeability. To meet this requirement, the inductance $L$ is constructed from the kinetic inductance of a tightly packed chain of $N \approx 10^2$ moderate-area ($\approx 1~\mu m^2$) Josephson tunnel junctions. One can interpret fluxonium as a transmon where the weak junction is galvanically short-circuited at low frequencies and hence there is no sensitivity to offset-charges even with $E_J/E_C \sim 1$.  Consequently, there is no need for a large shunting capacitance and hence circuit anharmonicity can be large. One can also view fluxonium as a generalized $N$-junction flux qubit, where the first and second order coupling to flux noise is suppressed as $1/N$ and $1/N^2$, respectively, without significantly reducing the frequency-tuning range. The circuit Hamiltonian is~\cite{koch2009charging}
\begin{equation}
H = 4E_C n^2 +
\frac12 E_L \phi^2 - E_J\cos (\phi - \phi_{\textrm{ext}}),
\label{Eq: Hamiltonian}
\end{equation}
where $\phi$ is the phase twist across the inductance and $2e\times n$ is the displacement charge at the capacitance. The two operators obey $[\phi, n] = i$. The quantity $\phi_{\textrm{ext}}$ is the reduced magnetic flux biasing the loop formed by the weak junction and the shunting inductance. At $\phi_{\textrm{ext}} = 0$, the low-energy spectrum corresponds to plasma-like oscillations in the central Josephson well with frequencies and transition dipoles similar to that of a transmon. In a previous experiment, we showed that at $\phi_{\textrm{ext}} \sim \pi/2$, the $0-1$ transition dipole can be exponentially suppressed when $E_J/E_C \gtrsim 10$, which leads to metastable states with relatively high coherence times (a few microseconds) limited by the first-order coupling to flux noise~\cite{lin2018demonstration}.

Here we operate the qubit near $\phi_{\textrm{ext}} = \pi$ (the so-called ``sweet spot"), where the first order sensitivity to flux vanishes by symmetry. Now the $0$ and the $1$ states correspond to the tunnel splitting of the two-fold degenerate classical ground state, and the qubit frequency is much lower than that at $\phi_{\textrm{ext}} = 0$. The non-computational states are separated by a plasmon gap and they form an anharmonic spectrum with a rich selection rule structure. By slowing down qubits about tenfold, to frequencies near $500~\textrm{MHz}$, energy relaxation is expected to naturally slow down as well without the need for improving the materials. Yet, fast logical gates can be constructed with such slowed down qubits, e.g. by driving the higher-frequency transitions to non-computational states~\cite{nesterov2018microwave}. The role of such transitions in creating strong interactions can be illustrated by the fact that a $0.5~\textrm{GHz}$ fluxonium can dispersively shift a $5-10~\textrm{GHz}$ readout mode by an amount comparable to conventional qubits despite the extreme frequency detuning~\cite{manucharyan2012evidence, zhu2013circuit, lin2018demonstration}.

Fluxoniums reported in this work spectacularly break the old unwritten rule of circuit design: for best coherence, the number of Josephson junctions in a physical qubit must be minimized. On the one hand, this rule was motivated by the belief that the Al/AlOx/Al Josephson tunnel junction is likely the faultiest part of a qubit circuit and hence the probability for a fatal failure might quickly grow with the number of junctions. On the other hand, this rule severely limits the design options for creating novel quantum hardware. For example, the minimal number $N=3$ was deliberately chosen in creating the three-junction flux qubit~\cite{chiorescu2003coherent}, which resulted in an impractically high sensitivity to the external flux~\cite{yoshihara2006decoherence, orgiazzi2016flux}. Unfortunately, the lack of high coherence seen in previous experiments on fluxoniums~\cite{manucharyan2012evidence, kou2017fluxonium, kou2018simultaneous} and other multi-junction qubits~\cite{gladchenko2008superconducting, bell2012quantum, bell2014protected} further contributed to discouraging the explorations of complex circuits, even though the possibly mundane origin of decoherence in those experiments was hidden by the low number of tested devices.

This paper is organized as follows. In section II we describe the measurements of 8 fluxonium devices. In section III we discuss our results, summarized in the Table 1, in the context of the known decoherence mechanisms. In section IV we offer a perspective on utilizing high-coherence fluxoniums in quantum computing. Section V summarizes the work.

\begin{figure}
	\centering
	\includegraphics[width=\linewidth]{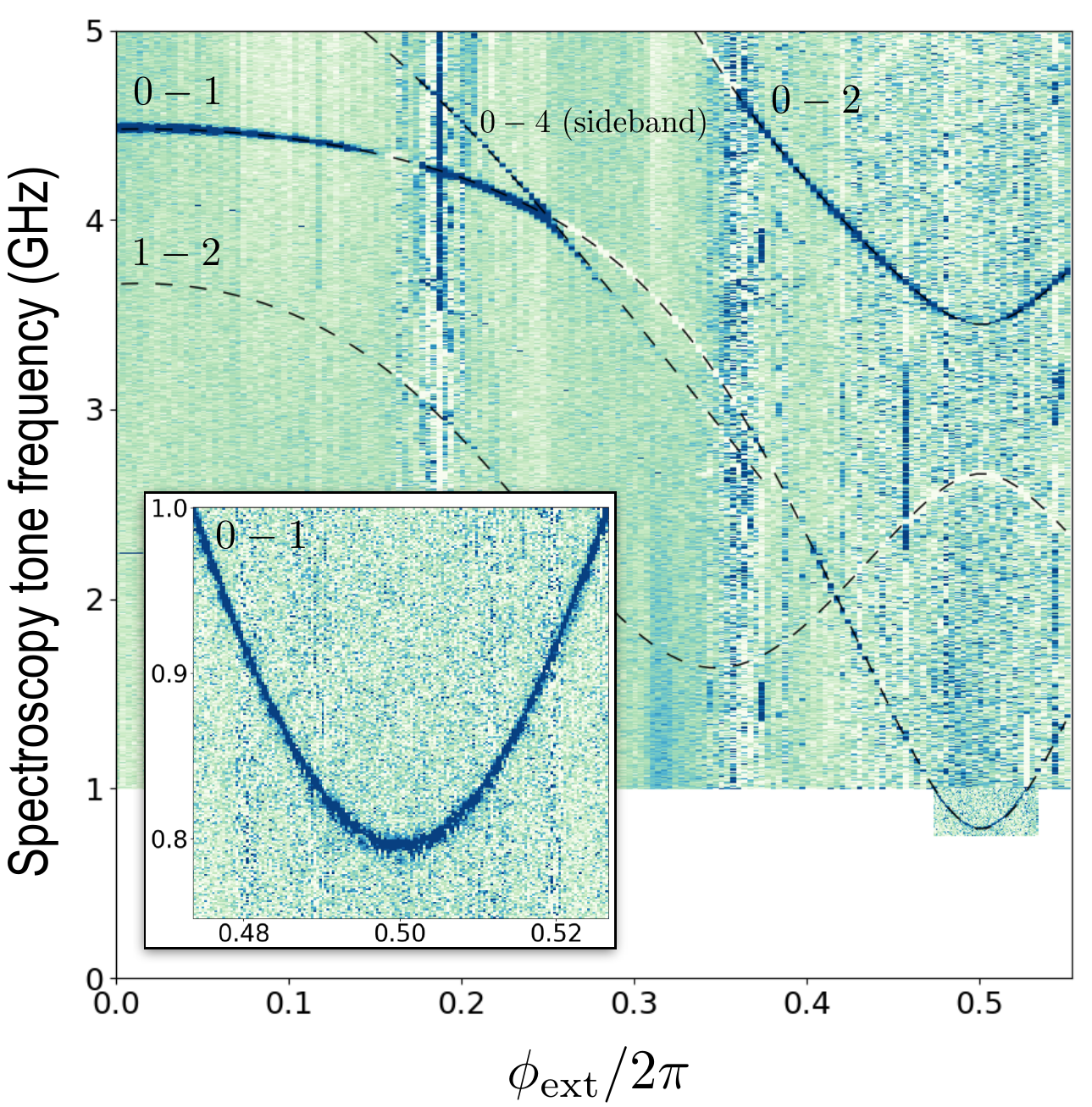}
	\caption{
		Two-tone spectroscopy transmission signal (arbitrary units) as a function of spectroscopy frequency and flux through the loop for device A. Lines indicate a fit to the spectrum of the Hamiltonian~(\ref{Eq: Hamiltonian}). The extra resonance line crossing (not anticrossing) the $0-1$ qubit transition is a red sideband of the $0-4$ transition and the readout at $7.5~\textrm{GHz}$. 
	}
	
	\label{fig:Fig2}
\end{figure}

\section{Experiment}

Experimental details on fluxonium devices used in this work can be found in Ref~\cite{lin2018demonstration}. Similarly to the original fluxonium design~\cite{Manucharyan113}, here we attach an external capacitance in the form of a simple dipolar antenna directly to the small junction. The capacitance in Fig.~\ref{fig:Fig1}a is mainly due to this antenna. The double-loop C-device is exactly the same used in Ref~\cite{lin2018demonstration} except it was measured at the simultaneous sweet spot (the so-called double sweet spot) of both loops, which were fabricated to have commensurate areas. In all other devices the split-junction was replaced by a single one for simplicity. The qubits were capacitively coupled to a 3D copper box readout mode with a frequency of $7.5~\textrm{GHz}$ and a  linewidth  $\kappa/2\pi  \approx 15~\textrm{MHz}$. The state of the qubit was monitored in a basic two-port cavity transmission measurement. For consistency, all devices were measured at a small external magnetic field. The precise effect of this field remains inconclusive at this stage.

Devices were fabricated in a single step using the four decade old Dolan bridge technique~\cite{dolan1977offset, frunzio2005fabrication}. This method is extremely robust for a typical junction area of about $1~\mu m^2$. It also yields the smallest possible stray capacitance in a planar geometry and up to $30,000-40,000$ junction long chains can be fabricated without a single fault. An image of a section of a typical fluxonium chain is shown in Fig.~\ref{fig:Fig1}b. The advantage of our chain design is that it does not require precision tuning of the resist mask. In fact the C-device was fabricated using low-voltage $20~\textrm{keV}$ electron beam lithography. For the substrate, we used high-resistivity silicon covered by the native oxide. No substrate surface treatment was applied apart from the low-power oxygen plasma ashing of the resist residues prior to the aluminum film deposition. The oxidation process yields conveniently high plasma frequencies in the vicinity of $20~\textrm{GHz}$.

The measured spectrum of the A-device together with the fitted theory lines is shown in Fig.~\ref{fig:Fig2}. The data was obtained by the conventional two-tone dispersive spectroscopy. Theory is a result of numerical diagonalization of the Hamiltonian~(\ref{Eq: Hamiltonian}) with the fit parameters being $E_J$, $E_C$, $E_L$, and the flux to coil current conversion. Note that in addition to fitting the two lowest frequency transitions, the theory precisely matches the red sideband of the readout mode with the $0-4$ transition. This indicates that there are no stray chain modes at frequencies below $10~\textrm{GHz}$ and the Hamiltonian~(\ref{Eq: Hamiltonian}) is an accurate model of our complex device. At $\phi_{\textrm{ext}} = 0$ the qubit transition corresponds to anharmonic oscillations in the central Josephson well. The nature of this transition, along with its frequency at about $4.5~\textrm{GHz}$ and the value of the transition dipole $\langle 0|\phi|1\rangle$ is similar to that of a typical transmon qubit. The difference here is that the Josephson well is slightly deformed by the shunting inductance (Fig.~\ref{fig:Fig1}c) and the anharmonicity, approximately given by $E_C \approx 0.8~\textrm{GHz}$, is considerably enhanced owing to the reduced value of the shunting capacitance. 

As the flux is tuned towards the sweet-spot at $\phi_{\textrm{ext}} = \pi$, the qubit transition monotonically drops to about $800~\textrm{MHz}$. Already with a naked eye it is evident that the sensitivity of the $0-1$ transition to flux does not exceed about $20~\textrm{GHz}$ per flux quantum, while the frequency is tuned by over two octaves. Due to the thermal occupation of the $1$-state one can see the transition $1-2$ in the small vicinity of $\phi_{\textrm{ext}} = \pi$. The transition $0-2$ is parity-forbidden exactly at $\phi_{\textrm{ext}} = \pi$, which is correctly reflected by the continuous reduction of its power-broadened linewidth (in a fixed power experiment) upon tuning the flux towards the sweet spot. The large anharmonicity of the qubit at the sweet spot can be characterized by the ratio $\omega_{12}/\omega_{01} \approx 3 - 10$ for our typical circuit parameters. Finally, we note that no two-level defects were spotted in the spectrum with anticrossings larger than a few $\textrm{MHz}$.


 \begin{figure}
 	\centering
 	\includegraphics[width=\linewidth]{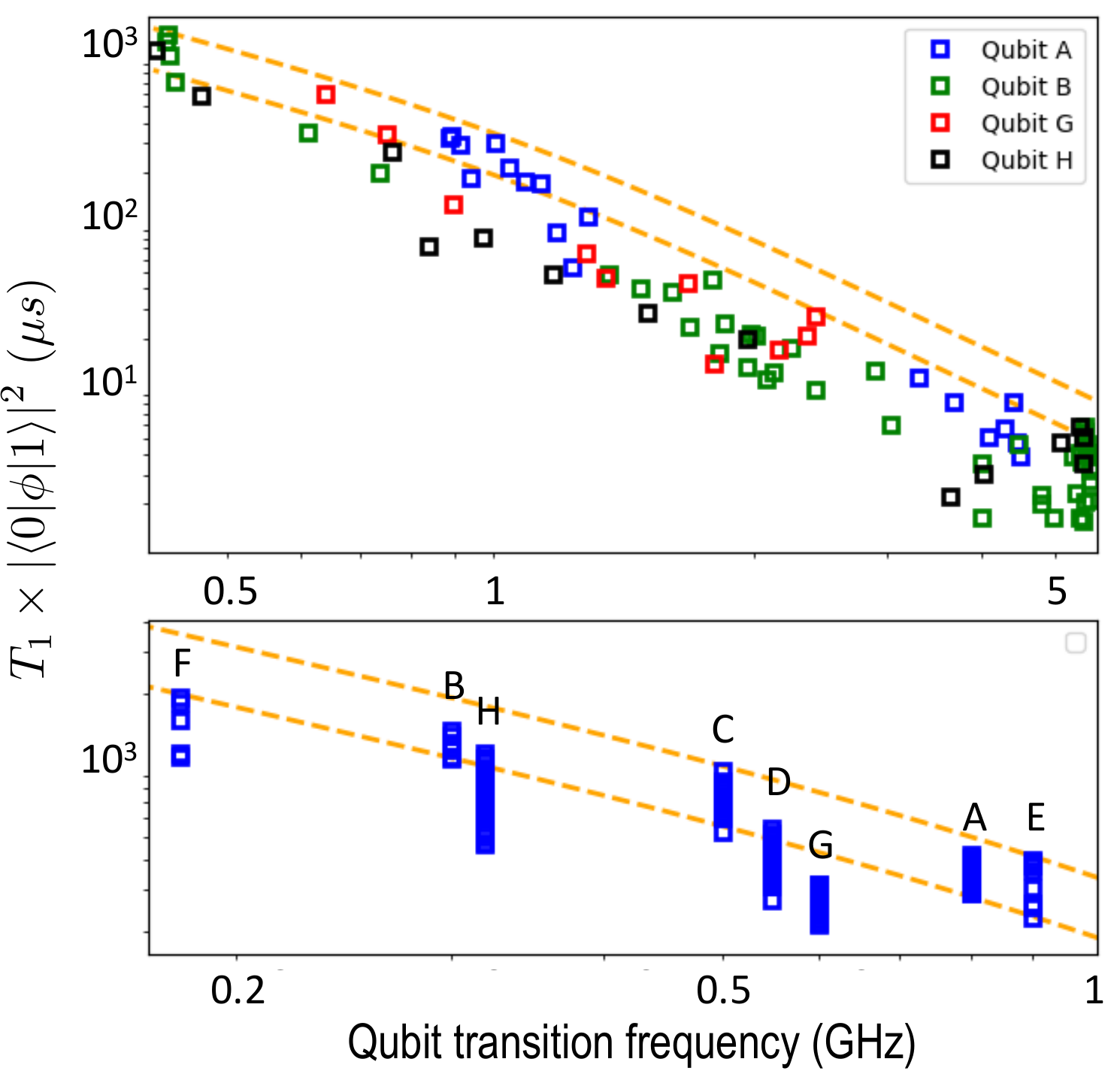}
 	\caption{
 		(top) Normalized energy relaxation time as a function of qubit transition frequency measured by tuning flux in devices A, B, G, H. (bottom) Same quantity, including repeated in time measurements, plotted for all devices biased at their half-integer flux sweet-spots. In both graphs, dashed lines represent a dielectric loss theory (see text) with $\tan\delta_C = (2.0-3.6)\times 10^{-6}$ at the frequency of $6~\textrm{GHz}$.
 	}
 	
 	\label{fig:Fig3}
 \end{figure} 

The frequency dependence of energy relaxation time $T_1$, covering several frequency octaves, was measured by flux-tuning the $0-1$ transition between $\phi_{\textrm{ext}} = 0$ and $\phi_{\textrm{ext}} = \pi$. The $T_1$ values were obtained by a standard time-domain experiment recording the evolution of the cavity transmission following a $\pi$-pulse to a qubit. The majority of relaxation signals fit well to a exponential function and the characteristic decay time is quoted as $T_1$. We did observe infrequent instances of a double-exponential decay; such data was discarded as being caused by a temporary instability in the system. 

\begin{figure*}
	\centering
	\includegraphics[width=\linewidth]{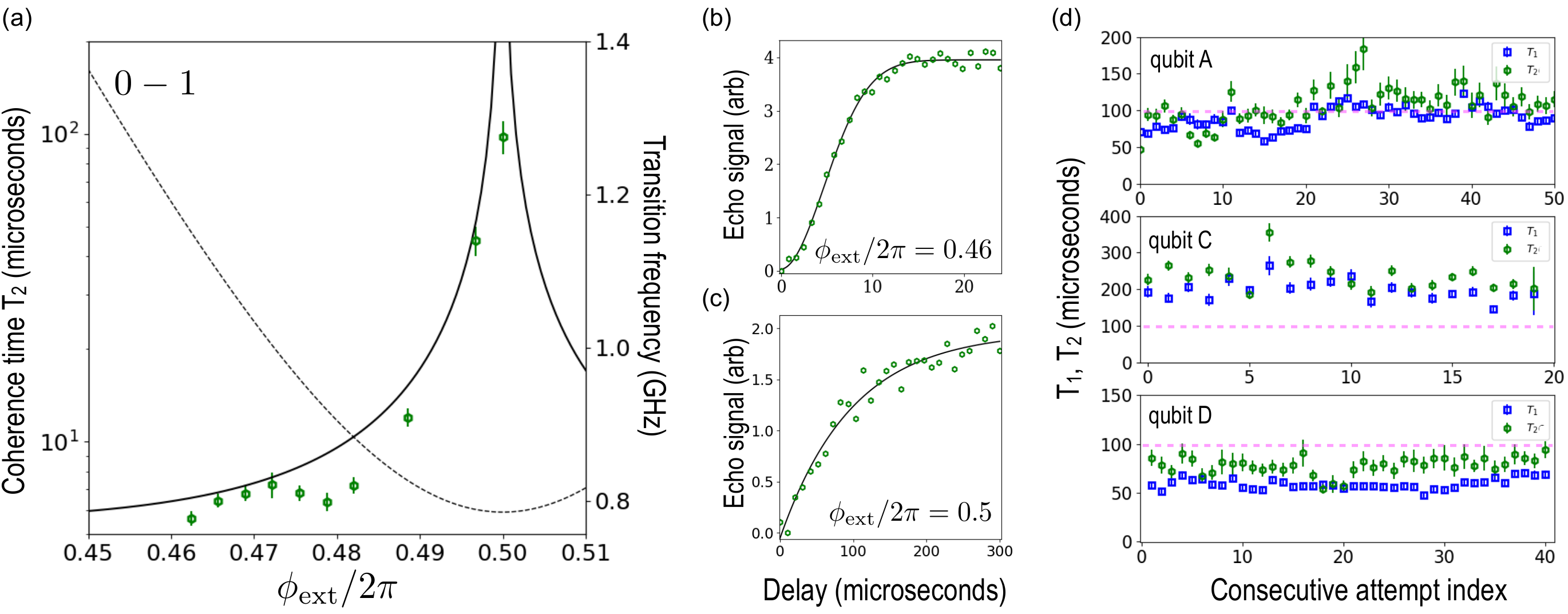}
	\caption{
		(a) Coherence time $T_2$ (markers) and the qubit frequency (dashed line) as a function of flux. Solid line indicates a prediction for the first-order coupling to a $1/f$ flux noise. (b) Gaussian echo signal far away from the sweet spot. (c) Exponential echo signal at the sweet spot. (d) interleaved measurement of temporal fluctuations of $T_1$ (blue markers) and $T_2$ (green markers) over a time interval of approximately 12 hours. 
	}
	
	\label{fig:Fig4}
\end{figure*} 

To extract the noise spectral density, we normalized the measured relaxation lifetimes by the square of the matrix element and plotted the quantity $T_1 \times |\langle 0| \phi|1\rangle|^2$ as a function of the qubit frequency $f_{01}$ (Fig.~\ref{fig:Fig3}). According to the Fermi's golden rule, this quantity is proportional to the spectral density $S_{\phi}(\omega)$ of the noise coupled to the phase variable $\phi$ at the qubit frequency $\omega$~\cite{clerk2010introduction}. At higher frequencies, the data, with some fluctuations, follows a simple dielectric loss model $S_{\phi}(\omega) = \hbar C\omega^2\tan\delta_C$, assuming an effective frequency-independent loss tangent $\tan \delta_C$ of the capacitance $C$. At lower frequencies it is necessary to take into account the stimulated emission due to the thermal occupation of the environmental modes. It is safe to assume that the temperature is at least $20~\textrm{mK}$, the approximate base temperature of our setup. This produces a notably \textit{lower} prediction for the values of $T_1$ than what was measured at low frequencies. In order to fix the discrepancy, we have to assume that the dielectric loss improves slightly towards lower frequencies through a phenomenological frequency dependence of the effective loss tangent, $\tan\delta \sim \omega^{\epsilon}, \epsilon>0$. Assuming $T > 20~\textrm{mK}$ we need $\epsilon \gtrsim 0.15$ to get a good agreement with the data, and a higher qubit temperature would require a larger value of $\epsilon$. This simple model is in agreement with the energy relaxation times of all devices measured at the sweet spot~(Fig.~\ref{fig:Fig3}, bottom). 


The dephasing measurement was performed using a standard single $\pi$-pulse echo sequence~\cite{yoshihara2006decoherence} in order to eliminate the uninformative low-frequency (minutes-scale) drifts in the setup. This protocol choice was also dictated by the relatively low readout efficiency in this particular setup, which required minutes of averaging time. The measured decoherence times $T_2$ as a function of flux for the A-device are shown in Fig.~\ref{fig:Fig4}. Away from the sweet spot, where the qubit transition is maximally sensitive to flux, we typically measure $T_2 \approx 3-6~\mu s \ll T_1$ and the echo signal has a distinct Gaussian shape. This confirms that decoherece is due to pure dephasing. Upon tuning the flux towards the sweet spot, we observe a monotonic rise in the coherence time to about $T_2 \approx 100~\mu s \gtrsim T_1$. Here the echo signal is exponential, which is consistent with relaxation-dominated decoherence. The flux-dependence of $T_2$ agrees with the prediction due to the first-order coupling to the $1/f$ flux noise with amplitude $2\times10^{-6}(h/2e)/\sqrt{\textrm{Hz}}$ at $1~\textrm{Hz}$.

To understand the robustness of the measured $T_2$ values at the sweet spot, we have performed repeated interleaved measurements of the times $T_2$ and $T_1$. The results for the representative devices A, C, and D are shown in  Fig.~\ref{fig:Fig4}d. Both the values of $T_1$ and $T_2$ fluctuate, but the fluctuations typically do not exceed a factor of two over about a half-day time interval. In some attempts we have observed $T_2 \approx 2 T_1$, but a typical situation is $T_1 <T_2 < 2 T_1$. We believe that the difference between $2T_1$ and $T_2$ may be caused by the combination of imperfect tuning of the echo pulses, the possible flux drifts (beyond the $1/f$ noise), and long averaging times. However, fluctuations in the pure dephasing time, formally defined as $1/(1/T_2 - 1/2T_1)$ could not be fully excluded at this stage. A setup involving a quantum-limited amplifier would clarify the details of the temporal fluctuations of coherence in future experiments. However, it is clear that the coherence is already largely limited by the energy relaxation, and hence the immediate next goal should be to further improve $T_1$. The longest reliably measured values of $T_2$ together with the accompanying value of $T_1$ are quoted for all devices in the Table 1. The D-device showed the lowest $T_2$ because of the noticeably lower $T_1$ value compared to the other devices. Interestingly, the longest coherence time ($T_2 > 300~\mu s$) was observed in the double loop C-device, where both loops needed to be simultaneously biased at the half-integer flux with a global magnetic field.

\section{Results and discussion}

\begin{table*}[t]
\centering
 \begin{tabular}{||c| c c c c c c c c c c c c c ||}
 \hline
 Qubit &$E_J$ & $E_C$ & $E_L$ & $N$ & $T_1$ & $T_2$ & $\omega_{01}/2\pi$ & $\omega_{12}/\omega_{01}$ & $\chi_{01}/2\pi$ & $\tan \delta_C$ & $\tan \delta_{\mathrm{AlOx}}$ &  $x$   &$\tan \delta_L$ \\ 
 \hline
  & GHz & GHz & GHz & - & $\mathrm{\mu s}$ & $\mathrm{\mu s}$ & GHz & - & MHz & $\times 10^{-6}$ & $\times 10^{-4}$ & $\times 10^{-8}$ & $\times 10^{-8}$\\
 \hline\hline
 A & 3 & 0.84 & 1 & 100 & 110 & 160 & 0.78& 3.4 & 0.27 & 1.7 &1.1 & 3.84  &  15.4\\ 
 \hline
 B & 4.86 &0.84 & 1.14 & 136 & 250 & 150 & 0.32& 11.1 & 0.57 & 1.5 &1.3& 0.52 & 2.03  \\
 \hline
 C & $2.2^*$ & 0.55 & 0.72 & 102 & 260 & 350 & 0.48& 3.8 & 0.08  & 1.15 &0.9& 1.77 & 5.75  \\
 \hline
 D & 2.2 & 0.83 & 0.52 & 196 & 70 & 90 & 0.56& 4.1 & 0.1 & 1.9 &4.0 & 7 &28.25 \\
 \hline
 E & 1.6 & 0.86 & 0.5 & 100 & 108 & 140 & 0.83& 2.5 & 0.05 & 3.25 &1.0& 7.8 & 30.22 \\
 \hline
 F & 3.4 & 0.8 & 0.41 & 348 & 270 & 165 & 0.17& 18.3 & 0.28  & 0.3 &4.5 & 0.63 & 2.1 \\
 \hline
 G & 1.65 & 1.14 & 0.19 & 400 & 110 & 140 & 0.55& 4.1 & 0.03 & 5.6 &3.8  & 8.65 & 34.9  \\
\hline
 H & 4.43 & 1 & 0.79 & 100 & 230 & 235 & 0.32& 11.8 & 0.1 & 1.1&0.9 & 0.72  & 2.85 \\
\hline
\end{tabular}
\caption{Summary of measured and inferred parameters of all fluxonium devices studied in this work. The quantities $E_J, E_C, E_L$ were obtained from spectroscopy fits; the qubit dispersive shift $\chi_{01}$ was calculated using data and formulas of Ref~\cite{lin2018demonstration}; the estimates of various loss mechanisms are discussed in Section III. 
$^*$effective $E_J$}
\label{table:qubit_summary}
\end{table*}

The summary of measured coherence times along with the extracted device parameters is given in Table 1. Using this information we can place important bounds on various decoherence mechanisms, which are summarized in the table as well and discussed in this section. 

\textit{Flux noise.} Dephasing times measured away from the sweet spot as a function of flux (Fig.~\ref{fig:Fig4}a,b) agree with a $1/f$ flux noise model with the amplitude $\approx 2\times 10^{-6}(h/2e)/\textrm{Hz}^{1/2}$ at $1~\textrm{Hz}$. Similar noise level was measured in flux qubits~\cite{orgiazzi2016flux}. However, here the off-sweet-spot coherence times are 2-3 orders of magnitude longer (a few microseconds), due to the large number of junctions $N$ and hence the proportionally reduced first-order flux sensitivity, whose maximal value is given by $2\pi E_L\times 2e/h \propto 1/N$~\cite{manucharyan2012evidence}. Having measured both the transition frequency vs. flux and the flux noise amplitude, we can estimate the limit on coherence time at the sweet-spot due to the second order coupling. Its value can be approximated by $(2\pi)^4 E_L^2/f_{01} \times (2e/h)^2$~\cite{manucharyan2012evidence}, which gives a range of coherence times $10-100~\textrm{ms}$ for the typical devices presented here. Such long times are possible entirely due to the $1/N^2$ scaling of the second-order flux sensitivity.

Interestingly, the data shows no signatures of flux noise induced energy relaxation, reported in recent experiments on flux qubits~\cite{yan2016flux, quintana2017observation}. Formally, the relaxation rate is given by the familiar Fermi's golden rule expression~\cite{clerk2010introduction} where the spectral density of the bath is replaced by that of the flux noise, which grows at low frequencies. In the case of a fluxonium, the relaxation rate scales as $E_L^2 \sim 1/N^2$ (the square of the energy matrix element in the Fermi's golden rule). This gives a protection against the suggested energy relaxation by a factor $10^2 - 10^4$ in comparison to flux qubits. Such a large protection factor may explain the qualitative difference between the $T_1$ data vs qubit frequency reported in ~\cite{yan2016flux, quintana2017observation} and that shown in Fig.~\ref{fig:Fig3} here.


\textit{Out-of-equilibrium quasiparticles.} A qubit can relax by emitting a photon which is absorbed by an unpaired quasiparticle tunneling across a junction~\cite{Martinis2009quasiparticle, Catelani2011quasiparticle2}. Assuming that the $T_1$ values in the Table 1 are limited by the tunneling of out-of-equilibrium quasiparticles across the chain junctions, we conclude that their normalized density $x$ can be as low as $x < 10^{-8}$. Given the device dimensions, this number corresponds to less than one quasiparticle in the entire chain. 

According to theory, the tunneling across the weak junction is coherently suppressed at $\phi_{\textrm{ext}} = \pi$ by the destructive electron-hole interference~\cite{Catelani2011quasiparticle1}. The interpretation of a previous fluxonium experiment in terms of this effect~\cite{pop2014coherent} implied a two orders of magnitude difference between the values of $x$ for the weak junction and for the chain junctions~\cite{vool2014non}, which seems unlikely. In our experiment, an estimate of the quasiparticles density near the small junction can be obtained from the double-loop C-device. There, the coherent suppression of tunneling is absent even at the double sweet spot of both loops because one of the two weak junctions is always away from the $\pi$-phase bias~\cite{lin2018demonstration}. Yet, the measured relaxation rate is similar to those of the single-loop devices. Its value requires the absence of quasiparticles near the weak junction, consistent with the above conclusion regarding the absence of quasiparticles in the chains. We cautiously speculate that vortices may be efficient at trapping quasiparticles in our specific device geometry~\cite{wang2014measurement}, although more experiments are needed to test this hypothesis.

\textit{Out-of-equilibrium photons.} Every qubit undergoing a dispersive readout will experience dephasing due to the photon shot noise if the readout mode is not properly thermalized~\cite{schuster2005ac}. In this work we deliberately avoided this issue by making the readout mode linewidth $\kappa$ much larger than the dispersive shift $\chi$ (Table 1). Of more importance is the photon shot noise in the $N-1$ collective modes of the Josephson chain~\cite{ferguson2013symmetries}. Their frequencies bunch near the junction's plasma frequency, which in our case is around $20~\textrm{GHz}$. Because of the non-linearity of the plasma modes, occupation of one such mode by a single photon introduces a dispersive shift of about $0.1\%$ of the qubit frequency. This shift is much larger than the qubit's natural linewidth~\cite{viola2015collective}. Hence, in order for the qubit to have a coherence time $T_2$, the average time for the absence of an out-of-equilibrium photon excitation in each mode must be longer than $N\times T_2$. Given numbers in Table 1, we estimate this time to be longer than $50~\textrm{ms}$, which means that the chain is practically empty of the out-of-equilibrium photons. The thermalization of plasma modes in our chains is intriguing because the microwave environment at such high frequencies is poorly characterized. Nevertheless, this behavior is consistent with the absence of quasiparticles in the chains, reported above.


\textit{Dielectric loss.} Because our devices have capacitive antennas, they are exposed to the surface loss similarly to any other capacitively-shunted qubit. The relaxation time $T_1$ grows upon reducing the qubit frequency in agreement with a model of a nearly frequency-independent loss tangent. To match the data at frequencies below $1~\textrm{GHz}$, a weak phenomenological frequency dependence ($\tan\delta_C \propto \omega^{\epsilon}$, $\epsilon \approx 0.15 - 0.5$ depending on the qubit temperature) of the loss tangent is required. However, because of the small value of $\epsilon$, this effect does not considerably influence our main conclusions. 
The values of $T_1$ at the sweet spot can be explained assuming a narrow range of $\tan\delta_C \approx (2.0-3.6)\times 10^{-6}$ taken at the frequency of $6~\textrm{GHz}$~(Fig.~\ref{fig:Fig3}). These numbers are considerably higher than the effective surface loss tangent reported for a number of transmon devices~\cite{wang2015surface}. Such a discrepancy is likely in part due to the sub-optimal aluminum on silicon fabrication process chosen for the present devices. Measurements in the transmon regime at $\phi_{\textrm{ext}} = 0$, indeed yield $T_1 \approx 5~\mu s$, which match to those of similarly fabricated Al on Si transmons with similar interface participation ratios~\cite{chu2016suspending}. Temporal fluctuations of $T_1$ in our devices may be consistent with the recent data on the X-mon qubits, explained by the drifts in the value of $\tan\delta_C$ due to the dynamics of the weakly-coupled two-level defects in the dielectric~\cite{klimov2018fluctuations}.

Dielectric loss in the chain junctions is another potentially important decoherence mechanism. Assuming that each junction's capacitance has a non-zero loss tangent, we can estimate its average value as $\tan\delta_{AlOx} < 10^{-4}$. Note, that the large number of junctions in the chain helps to reduce the relaxation rate proportionally to $1/N$. This is because the alternating voltage across the antenna is divided by $N$ for each junction of the chain. Our estimate on the loss tangent of AlOx is about an order of magnitude smaller than the previously reported bulk value of about $10^{-3}$~\cite{pappas2011two}. It is possible that the area of each chain junction (about $1~\mu m^2$) is sufficiently low to make encountering a strongly-coupled charge defect statistically unlikely. This effect can be explored in future devices upon varying the junction area.

\textit{Inductive loss tangent.} By analogy with the effective dielectric loss tangent $\tan \delta_C$ of the capacitance $C$ we can introduce a loss tangent $\tan\delta_L = \textrm{Im}[L]/\textrm{Re}[L]$ for the inductance circuit element. The convenience of this quantity is that the quality factor $Q$ of a resonator made out of such a lossy inductance and a perfect capacitance is given by $Q = \tan\delta_L^{-1}$. It is instructive to compare the two loss tangents $\tan\delta_L$ and $\tan \delta_C$ which would lead to the same value of $T_1$. One can show that $\tan\delta_L/\tan\delta_C = \omega^2 L C$. For our fluxoniums, $ \omega_{01}^2 L C \sim 10^{-1} - 10^{-2}$, which places a much more challenging requirement on the resonator quality factor test of the inductive element compared to the capacitive element. This is related to the fact that we need an unusually low quasiparticle density in the chains, $x < 10^{-8}$, in order to explain the measured $T_1$'s.

We conclude that the most likely explanation for the measured $T_1$ times in all devices is the surface loss in the antenna. By fabricating these devices either on sapphire ($Al_2O_3$) or on a properly surface-treated silicon, the $T_1$ times can be further increased by at least a factor of 3, which is sufficient to explore the limits to coherence, beyond the surface loss or flux noise, at the level of $1~\textrm{ms}$.

\section{Towards quantum computing with fluxoniums}
How can the high-coherence fluxonium qubits described above interact strongly on demand and undergo fast two-qubit gates? This question is especially relevant, because fluxoniums achieve their superior coherence largely due to the drastic reduction of the qubit transition frequency. Here we outline how fluxoniums can be integrated into all the existing schemes of scalable quantum computing. 

\subsection{Capacitive coupling }

Viewing the fluxonium circuit as an ``inductively-shunted charge qubit", one can understand the effect of connecting two such devices by a mutual capacitance $C_M$ using the charge qubits expressions. Assuming that $C_M \ll C$, the effective interaction term is given by $H_{\textrm{int}} = J_M n_1 n_2$, where $J_M =2 E_C\times (C_M/C)$ and $n_{1,2}$ are the charge operators of the two devices. Given that in our design $E_C \sim 1~\textrm{GHz}$, it is reasonable to count on $J_M \sim 100-200~\textrm{MHz}$, which are large even by the standards of conventional qubits.

\subsection{Inductive coupling }

Viewing the fluxonium circuit as a superconducting loop with a weak link, two such devices can be coupled extremely strongly by sharing one or several junctions between the two loops. Assuming that the fraction of the shared junctions is $m \ll 1$, one can show that the interaction term is given by $H_{\textrm{int}} = J_L (\phi_1/\pi) (\phi_2/\pi)$, where $J_L = m \pi^2 E_L$ and $\phi_{1,2}$ are the phase operators of the two devices. Normalization of the phase operators by $\pi$ is convenient because $\langle 0|\phi_{1,2}|1\rangle \approx \pi$ at the sweet spot. Therefore, even for a modest shared junction fraction $m = 0.1$ (about 10-20 junctions), we get $J_M \approx 0.5~\textrm{GHz}$. Note that such a value of exchange coupling is comparable to the qubit frequency. In fact, by making $m\sim 1$, a molecular-type strong binding of two fluxoniums has already been spectacularly demonstrated~\cite{kou2017fluxonium}.

\subsection{Dispersive cQED}

Circuit quantum electrodynamics with fluxonium qubits was described in~\cite{zhu2013circuit}. The most useful strong dispersive regime of cQED corresponds to $\chi \gg \kappa$. Introducing the dimensionless photon creation (annihilation) operator $a^{\dagger} (a)$, the coupling term is $H_{\textrm{int}} = g_C n i(a-a^{\dagger})$ for capacitive coupling and $H_{\textrm{int}} = g_L \phi (a+a^{\dagger})$ for inductive coupling. Given the discussion of inductive and capacitive coupling above, it is straightforward to achieve the values $g_C/2\pi, g_L/2\pi \approx 100~\textrm{MHz}$, typical of conventional qubits~\cite{kou2018simultaneous, lin2018demonstration}. However, here the qubit frequency (typically $500~\textrm{MHz}$)  is far detuned from the photon frequency (typically above $5~\textrm{GHz}$). Nevertheless, the shifts $\chi$ can be large because of the contributions of the transitions connecting either state 0 or state 1 to one of the non-computational states such that the transition frequency is near the readout mode~\cite{zhu2013circuit}.  In fact, the dispersive shift is non-zero even for a $\langle 0|\phi,n|1\rangle \rightarrow 0$~\cite{lin2018demonstration}. 


\subsection{Flux-controlled gates}
Since fluxonium's spectrum can be tuned by flux it is tempting to consider flux-controlled gate operations. Nanosecond-fast flux tuning requires a 2D cQED setup. Although our experiment was performed in a 3D setup, we used no features of the 3D that are likely to degrade coherence when moving to 2D. For instance, the readout mode is implemented by a copper cavity with a low loaded quality factor $Q \sim 500-1000$. The simplest gate is the analog of the C-phase gate for transmons, relying on the repulsion of the two-qubit states $11$ and $20$ or $02$~\cite{dicarlo2009demonstration}. The states repulsion can be generated by a direct capacitive or inductive connection of fluxoniums. In fact, one may expect an enhancement of gate fidelity because fluxoniums can maintain a relatively high coherence time $T_2 \approx 5~\mu s$ during the gate operation.

\subsection{Fixed-frequency qubit gates}

A recent proposal described a fast C-phase gate between two capacitively or inductively coupled fluxoniums obtained by applying a $\pi$-pulse to the transition $1-2$ of the target qubit, whose frequency shifts depending on the state of the control qubit~\cite{nesterov2018microwave}. In general, the quantum state leakage during such gate operations was shown to be remarkably low owing to the large anharmonicity of the non-computational part of the spectrum. Although microwave-activated gates still require static biasing of qubits, their advantage is that they are compatible with the 3D circuit QED architecture already demonstrated in this work.

\subsection{Quantum adiabatic optimization}

A network of interconnected fluxoniums, after the projection to the computational subspace, can implement a generic quantum spin-$1/2$ Hamiltonian:
\begin{equation}
H = \sum_{i,j} h_i^{Z} \sigma_{Z_i} + h_i^{X} \sigma_{X_i} + J^{XX}_{i,j} \sigma_{X_i}\sigma_{X_j}.
\label{Eq: ManyBody}
\end{equation}
Here the field $h_Z$ is the qubit transition frequency at the sweet spot, field $h_X$ is the detuning from the sweet spot, and $J_{XX}$ is the nearest neighbor coupling constant. The field $h_Z$ can also be tuned independently from $h_X$ by replacing a single weak junction by a split junction, as it was done in the C-device. Such a Hamiltonian is typically implemented using the system of semi-classical SQUID circuits to explore quantum annealing algorithms~\cite{johnson2011quantum}.

Our devices can provide a previously unavailable realization of this model.  (i) Owing to the extremely large anharmonicity, $\omega_{12}/\omega_{01} \gg 1$, a network of fluxoniums remains in its computational subspace even in the presence of multiple spin-flips, i.e. projecting to the computational subspace remains valid. (ii) Second, even far away from the sweet spot, we get $T_2 \sim 5 ~\mu s$, which translates into the level broadening of  about $30~\textrm{kHz}$. This number in principle allows resolving the many-body level spacing in a system of $10$ locally coupled spins. (iii) Finally, inductive connection readily allows a local coupling to multiple neighbors with the condition $J_{XX} \sim h_Z$.  The three conditions are simultaneously required for exploring the most intriguing scenarios of quantum many-body physics of spin systems. Fluxoniums are therefore well positioned for constructing the next generation of quantum annealers operating in a  highly-coherent regime where quantum speed up is expected from theory~\cite{boixo2016computational}
\subsection{Optimal qubit frequency}

It is interesting to discuss the choice of the optimal qubit frequency as the design, in principle, allows to reduce it to an arbitrary low value. We believe that the presented qubit frequency range around $500~\textrm{MHz}$ is currently the optimum for a number of reasons. 

First concern is the finite temperature of the qubit. Already at $500~\textrm{MHz}$, which translates to a temperature of $25~\textrm{mK}$, a significant population of state $1$ is expected. In principle, this is not a problem for a quantum processor as long as the energy relaxation time $T_1$ is sufficiently long. The qubits anyway need to be initialized with a high-fidelity, which can be done most reliably with a fast single-shot readout. However, it is convenient to be able to characterize low-frequency devices without the need to do so, and hence keeping the qubit frequency not far below the temperature is advantageous. More importantly, for $\hbar \omega \ll k_B T$, the relaxation time $T_1$ must be rescaled compared to its zero-temperature value due to the stimulated emission factor $T_1 \approx T_1 (T=0)\times (k_B T/\hbar \omega+1)$. We remark that the stimulated emission due to a non-zero temperature can substantially degrade the lifetime of imperfectly-protected topological qubits operating at a near zero transition frequencies. Another issue with going too low in frequency is the narrowing of the sweet spot, which would enhance the sensitivity of the system to drifts in the flux bias. 

From the technical viewpoint, the chosen frequency range appears particularly convenient for scaling: there is room to frequency-resolve neighboring qubits by spreading them by a few hundred $\textrm{MHz}$; Rabi-driving with a frequency up to $50~\textrm{MHz}$ can be applied even within the 3D circuit QED, which can provide $10~\textrm{ns}$-fast single-qubit pulses; cross-talks are in general expected to be reduced at lower frequencies; last but not least, qubit pulses can be done using cheaper digital electronics which can significantly reduce the cost per channel.  

\section{Summary}

We presented a specific design of fluxonium qubits which repeatedly yielded high coherence times, up to $T_2 > 300~\mu s$ at the half-integer flux bias. Compared to a typical flux qubit, the effect of the $1/f$ flux noise is practically eliminated by the large loop inductance ($L \sim 10^2~\textrm{nH}$) of the Josephson chain. Moreover, the qubit transition frequency can still be flux-tuned by many octaves while keeping the coherence time above a few \textit{microseconds} at an arbitrary flux bias, limited by the first-order coupling to flux noise. 

The presented fluxoniums are compatible with the transmon-based scaling architectures, which require connecting the coupling antenna directly to the small junction. This connection comes at the price of energy relaxation induced by the surface loss in the antenna. The surface loss was largely (but not completely) bypassed by reducing the qubit frequency by a factor of about ten, to a range around $\omega_{01}/2\pi \approx 500~\textrm{MHz}$.  Because the spectral density of the noise associated with the surface loss drops rapidly with frequency, the relaxation time of our qubits exceeded that of the best capacitively-shunted circuits despite rather sub-optimal fabrication. By upgrading the fabrication procedures to the state of the art, we expect to extend the coherence time to the range $T_2 > 1~\textrm{ms}$. 

Importantly, the low qubit frequency does not prevent fast gate operations or strong interactions in general. This is because the spectrum of a typical fluxonium is highly distinct from that of a weakly anharmonic oscillator and in fact it is reminiscent of atomic clocks. Transitions to the non-computational subspace belong to a conventional frequency and transition dipole range, which, as we outlined in section IV, can be utilized for creating fast flux- or microwave- activated quantum gates. Moreover, the inductive connection of fluxoniums via shared junctions can make the exchange coupling comparable to the qubit frequency without leaving the computational subspace. These large exchange couplings combined with the exceptional coherence times can be especially useful for constructing coherent quantum annealers. The next steps towards quantum computing with fluxoniums consist of demonstrating the high coherence in combination with a fast single-shot readout, a fast flux tuning, and a high-fidelity two-qubit gates.

In closing, let us remark on our specific implementation of large inductances required by a fluxonium using a Josephson chain. Following the viewpoint of minimizing the number of junctions per qubit, one may be tempted to replace the discrete chain by a patterned film of a highly-disordered superconductor with a comparable kinetic inductance. It is important to realize that the effective loss tangent of the inductance must be in the $10^{-8}$ range in order to reach the coherence times reported in this experiment. Whether such a low loss can be reached with dirty superconductors is an interesting question~\cite{hazard2018nanowire,zhang2018microresonators, grunhaupt2018granular,shearrow2018atomic,niepce2018high}. Our specific Josephson tunnel junction chain design (Fig.~\ref{fig:Fig1}b) was primarily motivated by the maximal simplicity of the qubit fabrication procedure: it takes place in a single step and does not require high-contrast lithography. The long coherence was made possible, in part, by the remarkably good thermalization of both the quasiparticles and the collective mode excitations in the chains. Understanding this effect in future experiments may have a big impact on quantum circuit design. In the meantime, our experiment demonstrated, for the first time, that a qubit's coherence time can be extended beyond the state of the art by increasing circuit complexity. 

We would like to acknowledge fruitful conversations with Maxim Vavilov, Chen Wang, Benjamin Huard, Ivan Pechenezhskiy, and Konstantin Nesterov. This work was supported by the Sloan Research Fellowship, NSF Career, NSF via PFC at JQI, and ARO-LPS. 

\appendix*

\section{Supplementary Information}

\subsection{Circuit QED}
Qubit-cavity vacuum coupling coefficient $g / 2\pi\approx 70\mathrm{MHz}$ and cavity's coupling to the readout $\kappa / 2\pi\approx 15\mathrm{MHz}$ are the same for A-F. G and H have different antennae, resulting in $g / 2\pi \approx 40\mathrm{MHz}$.
\subsection{Flux noise}
An inherent decoherence source in solid state devices is 1/f flux noise, found to originate from surface defects on the substrate. Its associated spectral density,
\begin{equation}
S_\Phi(\omega) = 2\pi \frac{A^2}{\omega},
\end{equation}
is found to affect superconducting flux qubit across more than an order of magnitude in frequency \cite{yan2016flux}. Measurements utilizing flux-sensitive devices found the noise amplitude to vary between $10^{-5}-10^{-6}\Phi_o$ \cite{yan2016flux}.

Away from the sweet spot, the qubit is sensitive to first order flux noise and it dephases following a gaussian function with rate

\begin{equation}
\label{eqn:gaussian_flux}
\Gamma = \frac{\partial \omega}{\partial \Phi}A\sqrt{\ln2}
\end{equation}

Measurement of $T_2$ away from sweetspot gives us the noise amplitude $A \approx 1.8\mu \Phi_o$. The large shunting inductance lowers the qubit's first order sensitivity, and we were able to achieve a few microseconds decoherence times, despite the flux noise amplitude similar to values reported previously.
\\
As the flux is tuned to half-integer flux, the qubit's first order flux sensitivity goes to zero. However, the second order sensitivity, $\partial^2 \omega / \partial \Phi^2$, reaches a local maxima, giving rise to the concern that flux noise can still limit qubit's coherence.
The relevant dephasing rate is \cite{ithier2005decoherence}
\begin{equation}
\label{eqn:flux_2ndorder}
\Gamma_{\Phi^2} = \frac{\partial^2\omega}{\partial \Phi^2}A^2
\end{equation}
Due to the low flux noise amplitude in our system, the second order effect is small, resulting in an upper $T_\phi$ limit $>10$ms for all qubits.\\

Another potential detriment  coming from flux noise at half integer fluxes is energy relaxation. The noise would couple to the qubit via the persistent current $\hat{I}_p=\hat{\Phi}/L$, resulting in relaxation rate \cite{quintana2017observation}
\begin{equation}
\label{flux_relaxation}
\Gamma_{ij}^\mathrm{\Phi} =\frac{1}{\hbar^2}\frac{1}{L^2}|\langle j|\hat{\Phi}|i\rangle|^2 S_\Phi(\omega_{ij})
\end{equation}
Again, due to the large inductance of the chain the relaxation time due to flux noise, increases as $L^2$. It would be in the order of hundreds of milliseconds, thus would not be a concern in fluxonium. This is quite remarkable because typical flux-tunable qubits' $T_1$ are limited at the low frequency sweet spot while fluxonium reaches maximum $T_1$ there. 

\subsection{Dielectric loss }
Dielectric loss in fluxonium can be modeled as a lossy shunting capacitor with admittance $Y_C(\omega)$\\
Using Fermi's golden rule \cite{Schoelkopf2003}, we can write the relaxation rate
\begin{equation}
\label{eqn:fermi_rule}
\Gamma_1(\omega_{01}) = \frac{1}{\hbar}|\langle0|\hat{\Phi}|1 \rangle|^2S_{\mathrm{diel}}(\omega_{01}),
\end{equation}
where the noise spectral density of a lossy capacitance is
\begin{equation}
\label{eqn:nsd_diel}
S_{\mathrm{diel}}(\omega_{01})=\hbar\omega\mathrm{Re}\left(Y_C(\omega_{01})\right)\left(\coth\left(\frac{\hbar\omega_{01}}{2k_BT} \right) +1\right),
\end{equation}
with 
\begin{equation}
\label{eqn:admittance_diel}
\mathrm{Re}(Y_C(\omega))=\frac{\omega C}{Q_\mathrm{diel}}.
\end{equation}
Here, $C$ is the effective capacitance, experimentally determined from $E_C = e^2/2C$, and $Q_\mathrm{diel}=\tan\delta_C^{-1}$ is the effective quality factor. 

Besides the large antenna whose dielectric loss was studied in details \cite{wang2015surface}, fluxonium consists of a long chain of Josephson junctions, which may contribute to $T_1$ limitation. There are two potential lossy effects from the chain, which we estimate below.

The chain junction has large areas, $0.4\mathrm{\mu m} \times 2\mathrm{\mu m}$ in this study, so the junction insulator may cause catastrophic dissipation, especially since amorphous AlOx is known to be lossy, with the loss tangent in the order or $10^{-3}$ \cite{pappas2011two}. However, from our measured relaxation time across several qubits, the loss tangent from AlOx in the junctions must have a much lower dielectric loss, as shown in table 1, with the associated loss tangent limit at $10^{-4}$.

The small spacing of the chain loop with $N$ junctions gives rise to a ground capacitance $C_g$, which collectively contribute to the effective capacitance of the qubit with $C_\mathrm{chain}=C_g N /6$. Nevertheless, even the longest chain with 400 junctions results in $C_\mathrm{chain}$ about a hundred times smaller than the effective capacitance of the whole circuit, so the normalization is minimal. A study on participation ratio of the dielectric loss coming from the chain would reveal further information on this loss channel.
\subsection{Inductive loss}
Analogous to dielectric loss, we can prescribe a lossy inductor $L\rightarrow L(1+\tan\delta_L)$ with admittance
\begin{equation}
    \mathrm{Re}(Y_L(\omega)) = \frac{\tan\delta_L}{\omega L}
\end{equation}
For the same relaxation time $T_1$, the ratio between inductive loss tangent and dielectric loss tangent would be  
\begin{equation}
    \frac{\tan\delta_L}{\tan\delta_C} = \frac{(\hbar \omega)^2}{8E_CE_L},
\end{equation}
which is much smaller than unity. 
\subsection{Quasiparticles}
Non-equilibrium quasiparticles have been shown to limit qubit's energy relaxation time $T_1$ and coherence time $T_2$. At the fluxonium's sweetspot, quasiparticle effect tunneling across the small junction is suppressed \cite{pop2014coherent}. However, if there are quasiparticles in the chain, they would still cause energy relaxation \cite{Catelani2011quasiparticle1}\cite{vool2014non} with rate
\begin{equation}
\label{eqn:qp_chain}
\Gamma_1(\omega_{01}) = |\langle0|\frac{\hat{\varphi}}{2}|1 \rangle|^2\frac{8 E_L}{\pi\hbar} x_{\mathrm{qp}} \sqrt{\frac{2\Delta}{\hbar\omega_{01}}}
\end{equation}
Our measured $T_1$ indicates that $x_{qp}$ is in the order of $10^{-8}$, corresponding to less than a single quasiparticle in the entire chain. We emphasize that this is the high limit for these quasiparticle, as we have yet observed any effect from them.
\subsection{Cavity temperature}
The thermal dephasing rate due to hot cavity photons follows \cite{gambetta2006photon}
\begin{equation}
\label{eqn:thermal_dephasing_exact}
\Gamma_{\mathrm{th}} = \frac{\kappa_{\mathrm{tot}}}{2}\mathrm{Re} \left[ \sqrt{\left(1+\frac{2i\chi}{\kappa_{\mathrm{tot}}} \right)^2 + \frac{8i\chi n_{\mathrm{th}}^{\mathrm{eff}}}{\kappa_{\mathrm{tot}}}} -1\right].
\end{equation}
The number of cavity thermal photons must be very high to limit the dephasing time to the order of a few hundred microseconds.\\

Even if the dispersive shift is enhanced by a hundred times to improve readout signal to noise, the low cavity thermal photons number achieved in other labs with similar cryogenic setup \cite{yan2016flux} would still result in more than a millisecond dephasing time limit.

\newpage
\bibliography{SuperconductingCircuits}

\end{document}